\documentclass[english,12pt,sort&compress]{iopart}
\usepackage[T1]{fontenc}
\usepackage[latin1]{inputenc}
\usepackage{float}
\usepackage{iopams}
\usepackage{setstack}
\usepackage{graphicx}
\usepackage{amssymb}
\usepackage[numbers]{natbib}
\usepackage{verbatim}

\newcommand{\prt}[0]{t}
\usepackage{babel}
\makeatother
\begin{document}

\title[]{Phase diagram of the ABC model with nonequal densities}

\author{O Cohen and D Mukamel}

\address{Department of Physics of Complex Systems, Weizmann Institute of Science, Rehovot 76100, Israel}

\ead{\mailto{or.cohen@weizmann.ac.il} and \mailto{david.mukamel@weizmann.ac.il}}

\begin{abstract}
The ABC model is a driven diffusive exclusion model, composed of three species of particles that
hop on a ring with local asymmetric rates.
In the weak asymmetry limit, where the asymmetry vanishes with the length of the system, the model
exhibits a phase transition between a homogenous state and a phase separated state.
We derive the exact solution for the density profiles of the three species
 in the hydrodynamic limit for arbitrary average densities.
The solution yields the complete phase diagram of the model and allows the study of the nature of
the first order phase transition found for average densities that 
deviate significantly from the equal densities point.

\end{abstract}


\pacs{05.50.+q, 05.70.Ln and 64.60.Cn}

\maketitle

\section{Introduction}

Systems that are driven out of equilibrium by an external field, such
as temperature gradient or electric field, have been studied extensively
in recent years.
In the absence of a general theory, insight into their properties can
be acquired by investigating simplified models.
Studies of numerous driven models involving some conserved quantity have shown that their steady state typically exhibits algebraic decay of
correlations \cite{Spohn1983,Garrido1990,Dorfman1994,Ortiz2004,Sadhu2011} and in some cases long-range order and symmetry breaking in one dimension \cite{Evans1995,Lahiri1997,Lahiri2000,Evans1998,Evans1998b}.
One particular model which has drawn much attention recently is the ABC model \cite{Evans1998,Evans1998b}. This
is a prototypical model for phase separation in one dimension.
It consists of a periodic lattice of length $L$ where each site is occupied by one of the three species of particles,  labeled $A,\, B$ and $C$. The model evolves by random sequential updates whereby particles on nearest neighbour sites are exchanged with rates
\begin{equation}
AB\overset{q}{\underset{1}{\rightleftarrows}}BA\,\qquad\,
BC\overset{q}{\underset{1}{\rightleftarrows}}CB\,\qquad\,
CA\overset{q}{\underset{1}{\rightleftarrows}}AC.\label{eq:ABCdynamics}
\end{equation}
While for $q=1$ the model relaxes to an equilibrium state with homogeneously
 distributed particles, it exhibits phase separation for any finite value of $q\neq 1$
in the limit of $L\to \infty$. Generically, for arbitrary choice of the number of particles of the three species, $N_A,N_B$ and $N_C$,
the model does not obey detailed balance and it relaxes to a nonequilibrium steady-state.
A unique feature of the ABC model is that in the special case of $N_A=N_B=N_C$ the dynamics
 obeys detailed balance with respect to an effective Hamiltonian with {\it long-range interactions} for arbitrary value of $q$.
 This Hamiltonian provides a rare opportunity to gain insight into the mechanism behind phase separation
  in one dimension induced by a drive in the bulk.

The ABC model has been considered in the weak asymmetry limit where
$q$ approaches $1$ in the thermodynamic limit as $q=\exp(-\beta/L)$ \cite{Clincy2003}.
When the rate of approach is faster than a critical value, namely for $\beta<\beta_c$, the model reaches a homogenous phase
in the limit of $L\to\infty$. For $\beta>\beta_c$ the model reaches an ordered phase with
three macroscopic domains, each predominantly occupied by one of the species.
The phase transition has been studied in the hydrodynamic limit by analyzing the linear response of the
homogenous phase to small inhomogeneous perturbations.
In the equal densities case, the transition was found to be continuous, taking place at $\beta=\beta_c=2\pi\sqrt{3}$.
The transition remains continuous for small enough deviation from the equal-densities condition,
 and becomes first order beyond a tricritical point at larger deviations.
Since the analysis was based on linear stability of the homogenous phase,
the full phase diagram of the model and the nature of the first order transition could not be explored.
This would require the knowledge of the density profiles of the three species in the ordered phase.

In the present paper we derive an exact expression for the steady-state density profiles
of the ABC model for arbitrary values of average densities and $\beta$ by solving the hydrodynamic equations corresponding
to the evolution of the model.
We use our results to investigate its phase diagram and the nature of the first order phase transition.
 Beyond the tricritical point we find a range of temperatures where both the homogeneous and ordered
 phases are locally stable.
 The phase to which the model eventually relaxes could in principle be determined by minimizing the
 large deviation function.
Since this function is known only in the limit of weak drive ($\beta \ll 1$) \cite{Clincy2003}
and for small deviations from the homogenous phase \cite{Bodineau2008}, we can only draw the stability limits
of each phase. These limits define the region of parameter space where both phases are locally stable.

The ABC model has recently been generalized to include particle-nonconserving processes and its phase diagram
has been analyzed in the equal densities case \cite{Lederhendler2010a,Lederhendler2010b}.
 The phase diagrams of the canonical (particle-conserving) and grand canonical
(particle-nonconserving) ensembles have been shown to be inequivalent. This is in accordance with what is generally
expected in equilibrium systems with long-range interactions.
The study presented in this paper of the nature of the phase-separated state can be generalized to the case of the nonconserving ABC
 model with arbitrary densities. This would enable one to explore phenomena such as inequivalence of ensembles in a genuinely driven model which does not obey detailed
balance \cite{Cohen2011}.

The paper is organized as follows. We first present a brief review of the ABC model and
previous studies of its phase diagram in \sref{sec:ABC}. We derive the
steady-state of the hydrodynamic equations of the model in \sref{sec:exact_solution}, and
express it explicitly in terms of elliptic integrals in \ref{sec:elliptic}.
 In \sref{sec:phase_diagrams}
we study the resulting phase diagram of the model and compare it with results from Monte Carlo simulations.
The low temperature (strong drive) behaviour of the solution is derived in \ref{sec:low_temp}.

\section{Phase diagram derived from stability analysis}
\label{sec:ABC}
In this section we present the ABC model and review its properties and phase diagram, obtained in previous studies
using stability analysis of the homogeneous phase.

In studying the ordered phase of the ABC model one notes that for $q<1$ the ordered phase is
such that the domains  are arranged clockwise as $AA\ldots ABB\ldots BCC\ldots C$, and
counterclockwise for $q>1$. Throughout this paper we consider $q<1$. The case
of $q>1$ is obtained by permutating for instance the labels of the $B$ and $C$
in a system where the drive is given by $q'=1/q<1$.

As a result of the dynamical asymmetry, the model generically reaches a nonequilibrium
steady state with nonvanishing currents of particles.
The current of, say, the $A$ particles is proportional to the rate at which they perform a full clockwise
trip minus the rate of the counter-clockwise trip, yielding
\begin{equation}
J_{A}\sim q^{N_{B}}-q^{N_{C}}. \label{eq:currents}
\end{equation}
The other currents are obtained by cyclic permutation of the $A,B$ and $C$ labels.
While these currents vanish in the thermodynamic limit for arbitrary average densities,
 in the special case where $N_{A}=N_{B}=N_{C}=L/3$, the currents also vanish for finite systems with arbitrary length.
 In this case the dynamics obeys detailed balance with respect to an effective
long-range Hamiltonian  given by
\begin{equation}
\mathcal{H}\left( {\boldsymbol \zeta }
\right)=\sum_{i=1}^{L}\sum_{k=1}^{L-1}\frac{k}{L}
\left(A_{i}B_{i+k}+B_{i}C_{i+k}+C_{i}A_{i+k}\right), \label{eq:trans_invariant_H}
\end{equation}
where ${\boldsymbol \zeta }=\left\{{ \zeta}_i\right\}_{i=1}^{L}$ denotes a microstate of
the system such that $\zeta_i=A,B$ or $C$. The operators in the Hamiltonian are defined as
\begin{eqnarray}
A_{i}= \left\{\begin{array}{ccc}
1 & \: & \zeta_{i}=A \\
0 & \: & else \end{array} \right. .
\end{eqnarray}
and similarly for $B_i$ and $C_i$.
The probability of
a microscopic configuration is given by $P\left({\boldsymbol \zeta}
\right)\propto q^{\mathcal{H}\left({\boldsymbol \zeta}
\right)}$. The Hamiltonian yields a super-extensive energy which scales as $E\sim L^2$ with the system size, typical of systems with
long-range interactions.

As mentioned in the introduction, the ABC model is often considered in the limit of weak asymmetry , $q=\exp\left(-\beta/L\right)$, where $\beta$ is regarded as the inverse temperature of the model \cite{Clincy2003}.
This rescaling of the drive with $L$ corresponds to the Kac prescription for the rescaling of
 the temperature in long-range interacting systems \cite{Kac1963}.
It amounts to an effective rescaling of the energy so it becomes linear with the system size,
thus comparable to the entropy, $S\sim L$.
Study of this limit for the equal densities case revealed a second order phase transition
at $\beta=2\pi\sqrt{3}$ from the homogeneous state, where entropy dominates, to the ordered state
which is dominated by the energy term.

The ABC model has also been studied on an interval, where zero flux boundary condition is considered \cite{Ayyer2009,Barton2010,Barton2011}.
In that case the model obeys detailed balance also for nonequal densities and its steady state
can be obtained using the same effective Hamiltonian (\ref{eq:trans_invariant_H}).
The steady-state density profiles of the three species in the phase separated state of this model has been
evaluated for arbitrary values of average densities \cite{Ayyer2009}.
In the special case of equal densities, the steady state of the model on an interval and that on a ring
are related by a trivial mapping, allowing us to use the studies of the model on the interval
as a point of reference for the present work.

In the case of equal densities on a ring or arbitrary densities on an interval, where an effective Hamiltonian
can be defined, it has been demonstrated that due to the weak anisotropy limit local density correlations
vanish for $L\to \infty$. Namely,
\begin{equation}
\label{eq:MF_app}
\left\langle X_iY_{i+1} \right\rangle = \left\langle X_i \right\rangle \left\langle Y_{i+1} \right\rangle + O(1/L).
\end{equation}
where $X,Y$ denote either $A,B$ or $C$ and $\left\langle \quad \right\rangle$
denotes an ensemble averages over the steady-state distribution.
It has been argued that this lack of local correlation is valid also for nonequal densities on a ring \cite{Ayyer2009}. As a result of \eref{eq:MF_app} the hydrodynamic equations
\cite{Evans2000,Fayolle2004,Ayyer2009} corresponding to this model are given by
\begin{eqnarray}
\label{eq:meanfieldA}
\frac{\partial\rho_\alpha}{\partial\tau} & = &  \beta\frac{\partial}{\partial x}\left[\rho_\alpha\left(\rho_{\alpha+1}-\rho_{\alpha+2}\right)\right]+ \frac{\partial^{2}\rho_\alpha}{\partial x^2},
\end{eqnarray}
where $\tau$ is the macroscopic time-scale and $\rho_\alpha(x)$
is the coarsed-grained density profile of particles of type $\alpha$ for $x\in [0,1]$.
 The index $\alpha$ denotes the species and runs cyclicly over $A\,,B$ and $C$. The conservation of particles
 implies that $\int_0^1dx\rho_\alpha(x)=N_\alpha/L\equiv r_\alpha$, where $r_\alpha$ is the average density of
 particle $\alpha$. Since $r_A+r_B+r_C=1$ it is convenient to express the densities
 in terms of two independent variables as
\begin{equation}
(r_A,r_B,r_C) = (\frac{1}{3},\frac{1}{3},\frac{1}{3}) + 2\Delta (\sin{\phi},\sin{(\phi+\frac{2\pi}{3})},\sin{(\phi+\frac{4\pi}{3})}),
\end{equation}
where  $\Delta^2=\frac{1}{6}\sum_{\alpha=A,B,C}(r_\alpha-\frac{1}{3})^2$ is a measure
for the deviation from equal densities and $\phi$ is a phase variable.

It is easy to see that the homogenous profile, $\rho_\alpha(x)= r_\alpha$, is a solution of
 \eref{eq:meanfieldA}.
 Its stability with respect to small anisotropic perturbations revealed a critical line given by
\begin{equation}
\beta=\frac{2\pi\sqrt{3}}{\sqrt{1-36\Delta^2}}.
\label{eq:beta_c}
\end{equation}
The homogenous phase was found to be unstable at temperatures ($T=1/\beta$) below this line \cite{Clincy2003}.
 Probing the region just below the critical
line, infinitesimal perturbations around the homogenous phase were found to be stable only when $r_\alpha$ obey
\begin{equation}
\label{eq:TCP}
S\left( r_A,r_B,r_C\right)=\left( r_A^2+r_B^2+r_C^2 \right) - 2 \left( r_A^3+r_B^3+r_C^3 \right) < 0.
\end{equation}
For these values the model undergoes a continuous second order transition at (\ref{eq:beta_c}),
whereas for other values of $r_\alpha$ the transition becomes first order.
The tricritical line, where the order of the transition changes,
 is given by $S\left(r_A,r_B,1-r_A-r_B\right)=0$ and in terms of $\Delta,\phi$ by
\begin{equation}
108\sin\left(3\phi\right)\Delta^3-54\Delta^2+1=0.
\end{equation}

\begin{figure}
\noindent
\begin{centering}\includegraphics[scale=0.6]{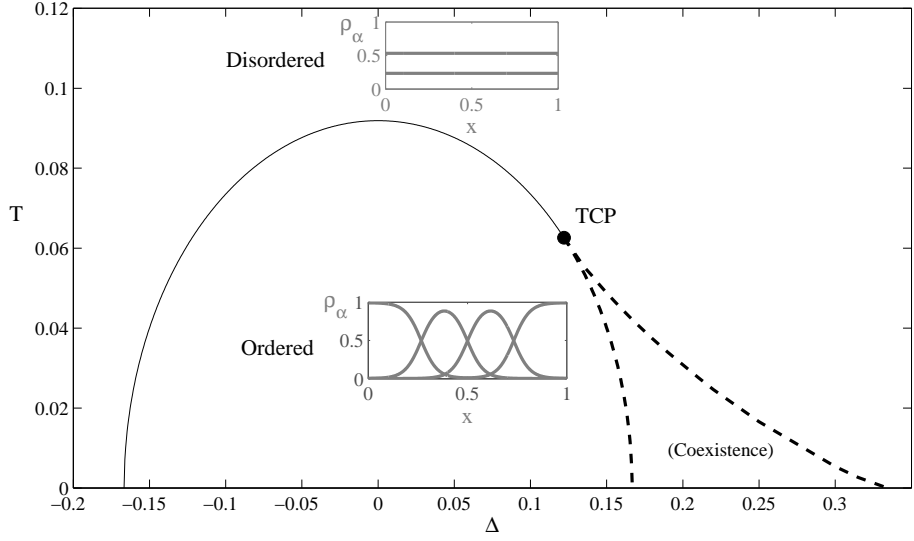}\par\end{centering}
\caption{
\label{fig:phase_diagram}
($T\equiv 1/\beta,\Delta$) phase diagram of the ABC model for two nonequal densities. The solid line
denotes the second order phase transition which turns into first order at the tricritical point (TCP).
The first order transition is depicted by two stability line (dashed lines) where the two phases
coexist.}
\end{figure}

The resulting phase diagram is shown in \fref{fig:phase_diagram} for the case of two nonequal densities, defined by taking $\phi=7\pi/6$ as
\begin{equation}
\label{eq:twoequal}
r_A=r_B=1/3-\Delta, \qquad r_C=1/3+2\Delta.
\end{equation}
The critical line and tricritical point in the figure are based on the work of Clincy et al., while
the upper stability line is drawn based on the results presented in the two following sections.
 Note that the phase diagram is not symmetric around $\Delta=0$. At one end of the phase diagram, for $\Delta=1/3$,
we obtain $r_C=1$ and hence no dynamics, whereas for $\Delta=-1/6$ we obtain the weakly asymmetric
exclusion process with $r_A=r_B=1/2$ \cite{Spohn1983}.

While the critical line and the tricritical point can be found by expanding \eref{eq:meanfieldA} near
the homogenous solution, studying the first order transition and the stability limit of the phase separated
state requires the knowledge of exact density profiles.
In this paper we calculate the steady-state density profiles of the model
 and use them to analyze its complete phase diagram.
An exact solution of \eref{eq:meanfieldA} on an interval
has been derived by Ayyer et al. \cite{Ayyer2009}.
 Following a similar derivation, we generalize their solution to the nonequal-densities regime of the
periodic model.
This allows us to study the nature of the first order transition phase predicted by Clincy et al.

\section{Density profiles for nonequal densities}

\label{sec:exact_solution}

In this section we study the steady-state solutions of the hydrodynamic equations of the ABC model (\ref{eq:meanfieldA}).
Assuming time-independent solutions, we omit the time derivative from \eref{eq:meanfieldA}
and integrate over $x$ to yield
\begin{eqnarray}
\label{eq:IntegratedMF}
\frac{\partial\rho_{\alpha}}{\partial x}=
-\beta \left[\rho_{\alpha}\left(\rho_{\alpha+1}-\rho_{\alpha+2}\right)\right]-J_\alpha,
\end{eqnarray}
where the constants of integration, $J_\alpha$, are interpreted as the steady-state currents of particles. They can be shown to obey $J_A+J_B+J_C=0$.
In order to solve \eref{eq:IntegratedMF} we apply several transformations
 which are similar to those used by Ayyer et al.
 for the ABC model on an interval \cite{Ayyer2009}.
 Multiplying \eref{eq:IntegratedMF} by $\rho_{\alpha+1}\rho_{\alpha+2}$
 and summing the three resulting equations yields
\begin{eqnarray}
\frac{\partial}{\partial x}\left(\rho_{A}\rho_{B}\rho_{C}\right) & =& -J_{A}\,\rho_{B}\rho_{C}-J_{B}\,\rho_{C}\rho_{A}-J_{C}\,\rho_{A}\rho_{B} \nonumber \\
 & = &-\frac{J_{A}}{\beta}\frac{\partial\rho_{C}}{\partial x}+\frac{J_{C}}{\beta}\frac{\partial\rho_{A}}{\partial x}.
\end{eqnarray}
Integrating this equation over $x$ yields a simple relation between the density profiles,
\begin{equation}
\label{eq:KDef}
\rho_{A}\rho_{B}\rho_{C}=K-Q_{A}\,\rho_{C}+Q_{C}\,\rho_{A},
\end{equation}
where $Q_\alpha\equiv J_\alpha/\beta$ and $K$ is a constant of integration.
One can check that this equation is indeed
invariant under cyclic permutations of $A,B$ and $C$ up to a change in the constant of integration, $K$.
\Eref{eq:KDef} is a generalization of the relation obtained for
the equal densities case, where $\rho_A(x)\rho_B(x)\rho_C(x)$ has been shown to be
 constant in space \cite{Fayolle2004,Ayyer2009}.
Using \eref{eq:KDef} in conjunction with $\rho_B=1-\rho_A-\rho_C$
allows us to express $\rho_A$ in terms of $\rho_C$ as
\begin{equation}
\label{eq:rhoAofrhoC}
\rho_{A}=\frac{-\left(\rho_{C}^{2}-\rho_{C}+Q_{C}\right)\pm\sqrt{\left(\rho_{C}^{2}-\rho_{C}+Q_{C}\right)^{2}+4Q_{A}\rho_{C}^{2}-4K\rho_{C}}}{2\rho_{C}}
\end{equation}
Inserting this expression back in \eref{eq:IntegratedMF} for $\alpha=C$ yields an explicit equation for $\rho_C$,
\begin{equation}
\label{eq:decoupled_rhoC_3}
\frac{\partial\rho_{C}}{\partial x}=\pm \sqrt{\left(\rho_{C}^{2}-\rho_{C}+Q_{C}\right)^{2}+4Q_{A}\rho_{C}^{2}-4K\rho_{C}}.
\end{equation}
The plus and minus signs correspond to the two halves of the ring around the maximum of $\rho_C(x)$.
Taking the square of this equation and writing it in the rescaled
variables $\prt=2\beta x$ and $y(\prt)=\rho_C(x)$ we obtain
\begin{equation}
\label{eq:particleEOM}
\frac{1}{2}y'(\prt)^{2}+U_{K,Q_{A},Q_{C}}(y(\prt))=0,
\end{equation}
where
\begin{equation}
\label{eq:effictive_pot}
U_{K,Q_{A},Q_{C}}\left(y\right)=-\frac{1}{8}y^{2}(1-y)^{2}+\frac{2K+Q_{C}}{4}y-\frac{2Q_{A}+Q_{C}}{4}y^{2}-\frac{Q_{C}^{2}}{8}.
\end{equation}
\Eref{eq:particleEOM} can be viewed as an
equation of motion of a zero-energy particle with mass $1$ in a quartic potential.
\Eref{eq:particleEOM} and the derivation below can be written in terms of
 either of the three species by cyclic permutation of $A,B$ and $C$. In the nonequal densities case the
  quartic potential changes under this permutation, yielding a different profile for each species.

\begin{figure}
\noindent
\begin{centering}\includegraphics[scale=0.6]{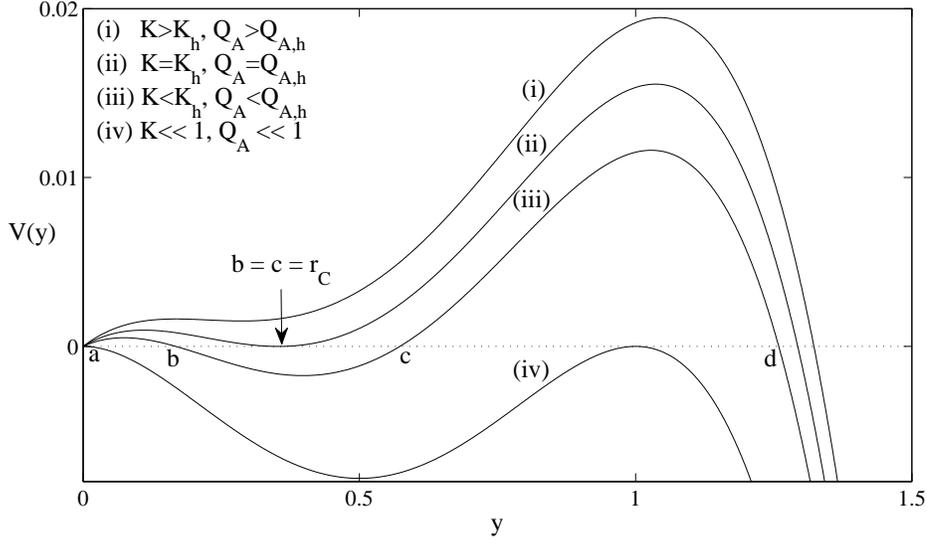}\par\end{centering}
\caption{
\label{fig:potential}
The effective potential, $U(y)$, for $r_A=r_B\neq r_C$ and some
values of $K,Q_A$ ($Q_C=0$). (i), (ii) and (iii)
represent the case where potential has 2,3 and 4 real roots. They
 correspond
to the case of the model has no physical solution, a homogenous solution and an
ordered solution, respectively.
The four roots of the potential in (iii) are denoted on the graph by $\{a,b,c,d\}$. (iv) depicts the limit of $K\ll 1, Q\ll 1$
which corresponds to the low temperature limit ($\beta \gg 1$) . }
\end{figure}

Depending on the values of $K,Q_A$ and $Q_C$ the potential may have two, three or
four real roots, depicted in the (i),(ii) and (iii) lines in \fref{fig:potential}, respectively.
 The four roots of the potential, denoted as $\{a,b,c,d\}$ can be shown
 to obey $0\leq a<b<c<1<d$. In this case the particle oscillates
 between $b$ and $c$ which is the only physical
trajectory. This is because we require that both $0\leq y(t) \leq 1$ and $U(y)\leq 0$.
 The case of three roots, when $b=c$, yields a constant trajectory in time which
  corresponds to the homogenous solution, $\rho_C(x)=y(2\beta x)=r_C$. The case
  where there are only two real roots does not correspond to any physical solution.
The manifold which defines the region of $\{K,Q_{A},Q_{C}\}$-space where the physical solution
resides is thus obtained by inserting the
 homogenous solution, $\rho_\alpha(x)=r_\alpha$, into \eref{eq:IntegratedMF} and (\ref{eq:KDef}) as
\begin{equation}
\label{eq:Kh_Qh}
Q_{\alpha,h}=r_\alpha \left(r_{\alpha+2}-r_{\alpha+1}\right), \qquad
K_h=r_A r_B r_C + Q_A\,r_C - Q_C\, r_A.
\end{equation}

The trajectory of \eref{eq:particleEOM} between $b$ and $c$ for $b<c$ yields the ordered
profile of the ABC model corresponding to given values of $K,Q_A$ and $Q_C$.
 In order to relate these parameters to the original parameters of the
 problem, $\beta$ and $r_\alpha$, we examine the period of oscillation of the particle between
$b$ and $c$, which we denote as $\Theta$.
 The periodic boundary condition of the profile, $\rho_C(x+1)=\rho_C(x)$, imposes a constraint on the solution
of the form, $y_m(t+2\beta)=y_m(t)$ or equivalently $\Theta=2\beta / m$.
 Here, the positive integer parameter $m$ corresponds to the
number of times the particle oscillates between $b$ and $c$ in a time interval of length $2\beta$.
We argue in the next section that only the $m=1$ solution describes the ordered steady state of the model.
The periodic boundary condition may be written as
\begin{equation}
\label{eq:integral_cond1}
1=\int_0^1dx=\int_0^{2\beta} \frac{d\prt}{2\beta}= \frac{m}{\beta}\int_{b}^{c}\frac{dy}{\sqrt{-2U_{K,Q_{A},Q_{C}}(y)}}.
\end{equation}
 An additional constraint on $y_m(t)$ comes from the total number of $C$ particles,
\begin{equation}
\label{eq:integral_cond2}
r_C=\int_0^1dx\rho_C(x)=\int_0^{2\beta} \frac{d\prt}{2\beta} y_m(\prt) =
\frac{m}{\beta}\int_{b}^{c}\frac{y dy}{\sqrt{-2U_{K,Q_{A},Q_{C}}(y)}}.
\end{equation}
The third constraint is obtained by dividing \eref{eq:IntegratedMF} by
$\rho_\alpha$ and integrating the result over $x$ using periodic boundary conditions.
For $\alpha=C$ the result yields the condition
\begin{equation}
\label{eq:integral_cond3}
 \frac{r_B-r_A}{Q_C} =\int_0^1dx\rho_C^{-1}(x)=\int_0^{2\beta} \frac{d\prt}{2\beta}  y_m^{-1}(\prt) =
\frac{m}{\beta}\int_{b}^{c}\frac{ y^{-1}dy}{\sqrt{-2U_{K,Q_{A},Q_{C}}(y)}},
\end{equation}
which is related to the difference between $r_A$ and $r_B$
and the consequent current of $C$ particles.
In \ref{sec:elliptic} we provide an analytic expression for \eref{eq:integral_cond1}-(\ref{eq:integral_cond3})
using elliptic integrals.

\begin{figure}
\noindent
\begin{centering}\includegraphics[scale=0.6]{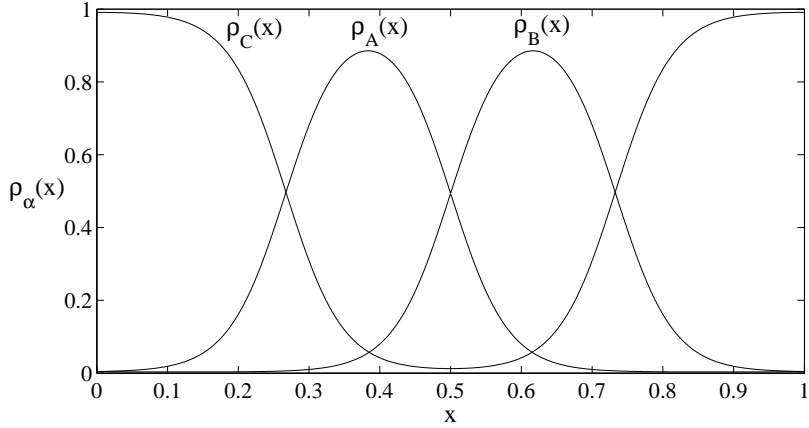}\par\end{centering}
\caption{
\label{fig:profile}
The ordered density profile obtained from \eref{eq:analytic_profile} and (\ref{eq:rhoAofrhoC})
for $r_A=r_B\neq r_C$ with $\Delta=0.1$, $\beta=25$ and $m=1$.
}
\end{figure}

 In the following section we will analyze the phase diagram which arises from the solution above.
 For simplicity we restrict ourselves to the two nonequal densities case, $r_A=r_B\neq r_C$, which
yields the same qualitative behaviour as the more general three nonequal densities case.
For  $r_A=r_B\neq r_C$ we find that $Q_C=J_C/\beta=0$.
 This simplifies the form of the effective potential (\ref{eq:effictive_pot}) and
leaves us with only two constraints, (\ref{eq:integral_cond1}) and (\ref{eq:integral_cond2}),
 which take the form of
\begin{eqnarray}
\label{eq:ellipctic_condition1}
\Theta=2\beta/m=2\varkappa K\left(1/k\right)/k, \\
\label{eq:ellipctic_condition2}
r_C=\frac{1}{3}+2\Delta=\frac{1}{\alpha_{-}}\left[ 2\frac{1+\alpha_-/\alpha_+}{\beta k /m\varkappa}
\Pi\left(\alpha_{-}^2/\alpha_{+}^2k^2 ,1/k\right)-1\right].
\end{eqnarray}
Here $\alpha_\pm,\varkappa$ and $k$ are functions of $K,Q_A$ given in \ref{sec:elliptic} and
$K(k),\Pi(n,k)$ denote the complete elliptic integral of the first and third kind, respectively, whose definition
is found in \ref{sec:elliptic} as well. The profile
of the $C$ particles is expressed by inverting the equation $x=\int_0^{\prt}d\prt '=\frac{1}{2\beta}\intop_{b}^{y}\frac{dy'}{\sqrt{-2U(y')}}$ as
\begin{equation}
\label{eq:analytic_profile}
\rho_C(x)=\frac{1+\mathrm{sn}\left(2\beta  x / \varkappa,k \right) }
{\alpha_+ - \alpha_- \mathrm{sn}\left(2\beta x / \varkappa,k \right)}.
\end{equation}
where $sn$ is the Jacobi's elliptic function \cite{Abramowitz64}. The dependence of the profile on $m$ is hidden in
the value of $K,Q_A$ which set $k,\varkappa,\alpha_\pm$.
The resulting $m=1$ profile for a specific value of $\beta$ is shown in \fref{fig:profile}.
In the section below we study the behaviour of this solution and
 the resulting phase diagram.
 We also examine its behaviour at low temperature ($1\ll \beta\ll L$) in \ref{sec:low_temp} and find that it
conforms with our physical understanding of the model.

It is interesting to note that the hydrodynamic equations of
the ABC model (\ref{eq:meanfieldA}) can be solved by considering a moving steady-state solutions of the form $\rho_\alpha\left(x,\tau \right)=\tilde{\rho}_\alpha\left(x+v\tau\right)$.
However, such solutions did not appear in the numerical relaxation of \eref{eq:meanfieldA}
as well as in Monte Carlo simulations.
 This may mean, although remains to be proven,
that moving solutions are unstable stationary solutions of the ABC dynamics.
We therefore restricted our derivation to case of $v=0$.

\section{Phase diagram for nonequal densities}
\label{sec:phase_diagrams}
In this section we examine the behaviour of $K,Q_A$ as we change the values of $\beta$ and $r_\alpha$ for the two nonequal densities case.
Their values are obtained by inverting the integral conditions given in \eref{eq:integral_cond1}-(\ref{eq:integral_cond2}),
which are written in an explicit form in \eref{eq:ellipctic_condition1}-(\ref{eq:ellipctic_condition2}).

\begin{figure}
\noindent
\begin{centering}\includegraphics[scale=0.6]{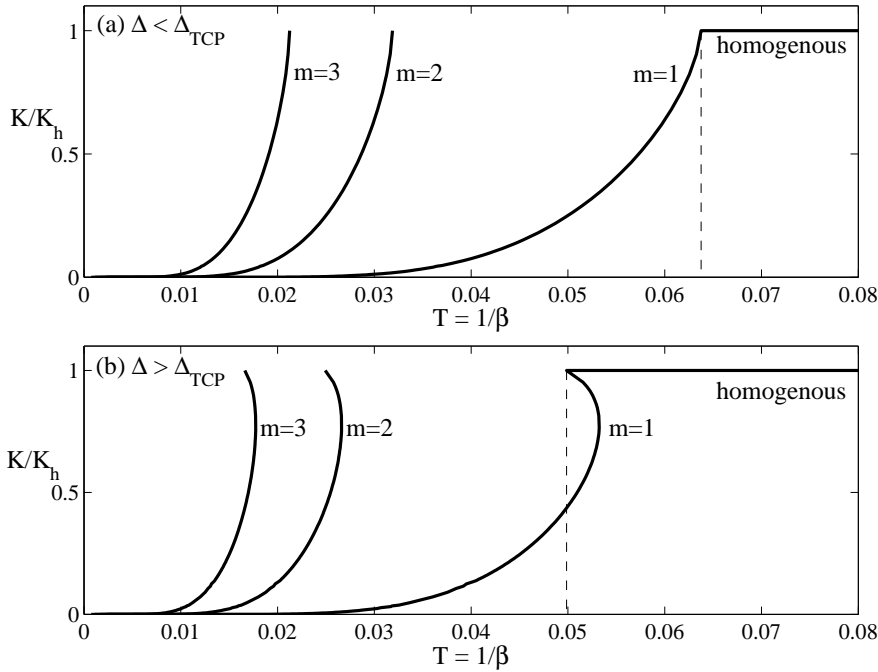}\par\end{centering}
\caption{
\label{fig:BetaK}
The temperature, $T\equiv 1/\beta=2/m\Theta(K,Q_A)$, computed as a function of $K$ for $m=1,2,3$ for
two nonequal densities, $r_A=rB\neq r_C$. $Q_A(K,\Delta)$ is set by \eref{eq:ellipctic_condition2}.
The dashed line denotes the critical point $\beta=2\pi\sqrt{3}/\sqrt{1-36\Delta^2}$. (a) and (b) are
calculated for $\Delta=0.12$ and $\Delta=0.14$, respectively, depicting the case
where $\Delta$ is below and above the tricritical point.}
\end{figure}

For all values of $\beta$ the hydrodynamic equations (\ref{eq:meanfieldA}) have a stationary
solution of the form $\rho_\alpha(x)=r_\alpha$, which corresponds to the homogenous values of
$K=K_h$ and $Q_\alpha=Q_{h,\alpha}$ given in \eref{eq:Kh_Qh}. As discussed in section \ref{sec:ABC}, this
solution becomes unstable below the critical line, $\beta>\beta_c=2\pi\sqrt{3}/\sqrt{1-36\Delta^2}$.
 In this regime we expect to find an ordered
solution. \Fref{fig:BetaK} displays $T=1/\beta$ computed according to \eref{eq:ellipctic_condition1}
where $Q_A$ is set for a given value of $K$ through \eref{eq:ellipctic_condition2}.
 For a small values of $\Delta$, in \fref{fig:BetaK}a,
we find a second order transition at $T=1/\beta_c$ between the homogenous phase and the $m=1$ ordered
 phase where $K<K_h$.
 This behaviour persists up to the tricritical point (\ref{eq:TCP}), which in the two nonequal densities case takes the simpler form of
$\Delta=1/\left(3+3\sqrt{3}\right)\simeq 0.122$. In \fref{fig:BetaK}b,
 we see that beyond the tricritical point, the $m=1$ ordered phase appears also at $T>1/\beta_c$. This is because the relation between
 $K$ and $\beta$ under a
fixed value of $\Delta$ is non-monotonic. As a result the model is expected to undergo a first order
transition between the two phases at value of $\beta$ between the two stability limits.
The discontinuity in $K$ at the transition implies that this is a transition from a homogenous state to an
ordered state with a finite amplitude of modulation.

In order to compute the first order transition point one has to know
the full large deviation function (LDF) of the ABC model, which is not known.
We may still draw the stability limits of the
two phases defined by the critical temperature and the minimum of
$K(\beta)$ in the ordered phase. The resulting stability lines are shown in \fref{fig:phase_diagram}
for the case of two nonequal densities.

\begin{figure}
\noindent
\begin{centering}\includegraphics[scale=0.6]{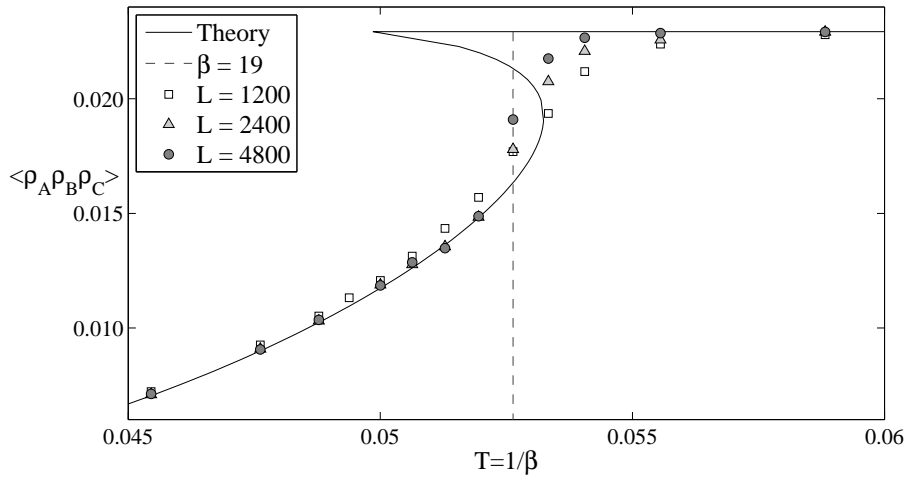}\par\end{centering}
\caption{
\label{fig:rABC}
The density of ABC triplets as measured in simulation as a function of $T\equiv 1/\beta$ in comparison
with the theoretical values (solid line). The simulation was performed for the two nonequal densities
case with $\Delta=0.14>\Delta_{TCP}\simeq 0.122$ and various values of $L$. The dashed line
denotes the value of $\beta=19$ for which the time evolution of $\langle\rho_A\rho_B\rho_C\rangle$ is plotted in \fref{fig:rABC_t}.
}
\end{figure}

In \fref{fig:rABC} we examine the first order transition using Monte Carlo simulations.
The algorithm for the simulation is straightforward.
At each step a site is selected at random and an exchange step is attempted where the particle in the chosen site
 may be exchanged with its neighbour to the right with probability given by
 \eref{eq:ABCdynamics}. We measured the parameter
$\left\langle \rho_A\rho_B\rho_C \right\rangle$ and compared it to that obtained from
the hydrodynamic solution by integrating \eref{eq:KDef} over $x$.
In the simulation it was measured by counting the number of $ABC$
triplets in the lattice after each $L$ exchange attempts and averaging the result
over many such time steps.

\begin{figure}
\noindent
\begin{centering}\includegraphics[scale=0.6]{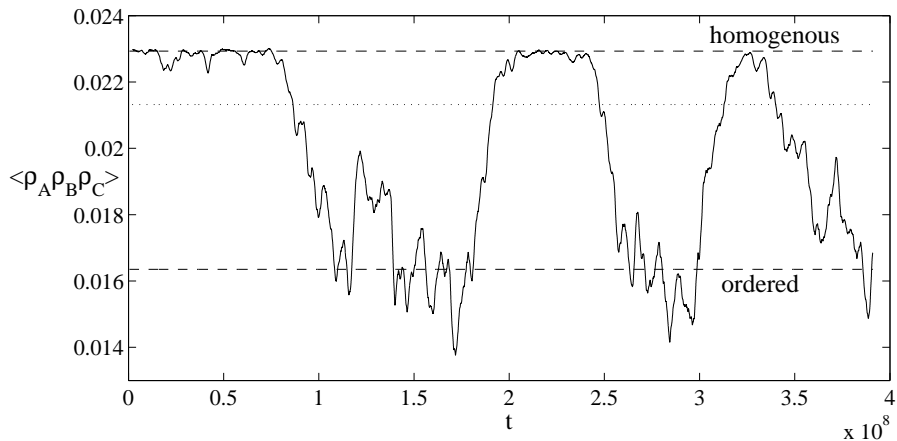}\par\end{centering}
\caption{
\label{fig:rABC_t}
The density of ABC triplets as a function of time given by the number of
Monte Carlo sweeps for $L=4800$, $\Delta=0.14$ and $\beta=19$. The upper and lower
dashed lines denote the theoretical values of the homogenous and ordered
phases, respectively. The dotted line denotes the unstable ordered solution which, as
expected, does not appear in the simulation.
}
\end{figure}

In \fref{fig:rABC} we plot the simulation results for different values of temperatures around the
 first order transition and for various system lengths.
 For each value of $T$ the simulation is initiated in the fully
ordered phase and run for a time period which was sufficient to observe transitions between the two phases.
 The number of $ABC$ triplets is averaged over the entire second half of the simulation
where the system is unaffected by its initial state.
The figure displays a first order phase transition, smoothen by finite size effects.
 The transition occurs below the critical point as suggested
by our analysis. Near the transition point we observe slow fluctuations of the system between
the two phases, as depicted in \fref{fig:rABC_t}. This implies that
the \fref{fig:rABC} might contain some errors near the transition point due to insufficient sampling
time.
We do not expect the transition point in the $L\to \infty$ limit to
obey Maxwell's construction since the horizontal axis is not the conjugate variable of $T$.
 The latter can only be derived from full LDF of the model.
\Fref{fig:rABC} also displays a good agreement with the theoretical
values for $\langle \rho_A\rho_B\rho_C \rangle$ above and below
the transition point, which confirms the validity of the mean-field approximation (\ref{eq:MF_app}).

\Fref{fig:rABC_t} depicts the fluctuations of the system between the ordered and disordered phases
for $L=4800$ at a temperature close the first order transition point. The figure shows significant and long-lived
fluctuations around the ordered phase. A thorough investigation of their nature showed that they are not
related to any known meta-stable state of the model and that they decay as the size of the system is increased. We avoid, however,
using larger systems since they would require much longer simulation time to display transitions between the two phases.

In \fref{fig:BetaK} we find that for $1/2\beta_c<T<1/\beta_c$ the $m=1$ is the only
stable solution, whereas the $m>1$ profiles are unstable.
 At lower temperatures the latter become stationary states and may theoretically be the
ground state of the model.
In the equal-densities case, this possibility has been ruled out by showing that
 the $m=1$ profile has the lowest free energy for all $T<T_c$ \cite{Ayyer2009}.
For nonequal densities, a similar analysis would require the knowledge of the full LDF of the
ABC model. Here, however, the fact that the $m=1$ is the ground state of the model
can be argued by noting that the $m^{th}$ solution corresponds to an ordered state with
particles segregated into $3m$ domains.
Since lower temperatures (stronger drive) favour segregation, it is natural to assume
that the most segregated state, $m=1$, remains stable for $T<1/2\beta_c$.
We therefore consider it to be the steady-state solution of
the model throughout the ordered phase. This argument is supported by Monte Carlo simulations
of the model, where only the $m=1$ profile was observed below the transition point.

\section{Conclusions}

In this paper we have derived an exact expression for steady-state density profile of the ABC model on a ring.
The derivation is based on a hydrodynamic description of the model
 which has been suggested to be exact in the thermodynamic limit \cite{Ayyer2009}.
The solution is valid for the case where the average densities of each species are not equal
and the model is thus out of equilibrium.
Using this solution we have studied the first order phase transition whose existence was suggested by
Clincy et al. The transition is of first order since there is a finite range of temperatures
where the model has two stable phases.
The transition point is located where the large deviation function of the two phases is equal.
However, since this function has not yet been found we can only draw
the stability limits of the two phases which define the coexistence region.
Monte Carlo simulation of a specific point in
parameter space confirmed that the first ordered phase transition occurs within the coexistence region,
above the previously known critical temperature.
The simulations also yielded a
good agreement with the hydrodynamic solution in regions where only one phase is stable.

The present study opens the door for future studies of the ordered phase
in the ABC model with arbitrary values of average densities. We plan to employ the solution obtained here
in order to study the inequivalence of ensembles in the nonconserving ABC model
with nonequal densities \cite{Cohen2011} and compare the results with those previously obtained
for the equal-densities case \cite{Lederhendler2010a,Lederhendler2010b}.


\ack We thank Amir Bar, Martin R. Evans and Ori Hirschberg for helpful
discussions. The support of the Israel Science Foundation (ISF) is gratefully acknowledged.

\appendix

\section{Analytic expression of the mean-field solution}
\label{sec:elliptic}
In this appendix we obtain an analytic expression for the relation between the parameters
of the effective potential $K,Q_A,Q_C$ and the parameters of the model, $\beta,r_\alpha$. We derive
the explicit expression for the general case and then present its simpler form for the two nonequal
densities case, $r_A=r_B\neq r_C$, mostly used in this study.

\subsection{Three nonequal densities}

In \sref{sec:exact_solution} we mapped the mean-field dynamics of the ABC model to
the motion of a particle in a quartic potential,
\begin{equation}
U_{K,Q_{A},Q_{C}}\left(y\right)=-\frac{1}{8}y^{2}(1-y)^{2}+\frac{2K+Q_{C}}{4}y-\frac{2Q_{A}+Q_{C}}{4}y^{2}-\frac{Q_{C}^{2}}{8}.
\end{equation}
The parameters of the potential $K,Q_A,Q_C$ are linked to the parameters of the
model, $\beta$ and $r_\alpha$, through three conditions (\ref{eq:integral_cond1})-(\ref{eq:integral_cond3})
which involve integration over the trajectory of the particle.

We begin with  \eref{eq:integral_cond1}
which can be expressed in terms of the incomplete elliptic integral of the first kind defined here
in the Jacobi form \cite{Abramowitz64},
\begin{equation}
F\left(x,k\right)=\intop_{0}^{x}\frac{dz}{\sqrt{\left(1-z^2\right)
\left(1-k^2z^2\right)}}.
\label{eq:elliptic_F}
\end{equation}
Following a similar derivation as in \cite{Ayyer2009} we introduce a M{\"o}bius transformation that
maps the roots of the potential,
\begin{equation}
U_{K,Q_{A},Q_{C}}\left( y \right) =0 \qquad \forall y\in\left\{a,b,c,d\right\},
\end{equation}
 onto the poles of the integrand in \eref{eq:elliptic_F},
\begin{equation}
\left\{a,b,c,d\right\} \mapsto \left\{-1,-1/k,1/k,1\right\}.
\end{equation}
The transformation is given by
\begin{eqnarray}
z=f\left(y\right)= \frac{d-a}{d+a} \frac{\alpha_+y-1}{ \left( \alpha_- - \frac{2ad}{d+a}\alpha_+\alpha_-\right)
 y+\left( 1 + \frac{2ad}{d+a}\alpha_-\right)}\equiv \frac{A y +B}{C y +D},
\end{eqnarray}
where
\begin{equation}
\alpha_	\pm = \frac{ \pm (b c - a d) +\sqrt{(a - b) (a - c) (b - d) (c - d)}}{ b c (a + d) - a d (b + c)},
\end{equation}
and
\begin{equation}
k=\frac{1+\alpha_- \left( b-a\frac{ (b-c) (b-d)}{a b-2 a d+c d} \right) }
{1- \alpha_+ \left(b-a\frac{ (b-c) (b-d)}{a b-2 a d+c d} \right)}
\end{equation}
The parameters $\alpha_{-}$,$\alpha_{+},A,B,C,D$ and $k$ are functions of $K,Q_A,Q_C$ through $a,b,c$ and $d$.
Let $\prt(y)$ be the time it takes the particle to move from $b$ to $y$. Using the transformation above
it may be expressed as
\begin{eqnarray}
\label{eq:HyberbplicTime}
\prt &= &2\intop_{a}^{y}\frac{dy'}{\sqrt{-2U(y')}} =  \varkappa \intop^{f(y)}_{-1/k}\frac{dz}{\sqrt{(1-z^2)(1-k^2z^2)}} \nonumber \\
& = &
\varkappa \left[ F\left(1/k,k\right)+F\left(f(y),k\right) \right],
\end{eqnarray}
where
\begin{equation}
\varkappa = \frac{2\sqrt{\left( C^2-A^2 \right)\left( C^2-k^2A^2\right)}}{AD-BC}
\end{equation}
\Eref{eq:integral_cond1}, whereby the period of oscillation obeys $\Theta=2\beta/m$,
can be expressed by setting $f(y)=1/k$ in
\eref{eq:HyberbplicTime}. For that end we first notice that the integral in this equation
can be brought to a simpler form in the new coordinates $w=k z$,
\begin{eqnarray}
\label{eq:complete_elliptic}
F\left(1/k,k\right)&=&\intop^{1/k}_{0}\frac{dz}{\sqrt{(1-z^2)(1-k^2z^2)}} \nonumber \\
 &=&\intop^{1}_{0}\frac{dw/k}{\sqrt{(1-w^2/k^2)(1-w^2)}}
=\frac{1}{k}K\left(1/k\right),
\end{eqnarray}
where $K\left(1/k\right)$ is the complete elliptic integral of the first kind. Using this form
the condition of $\Theta=2\beta/m$ can be written as
\begin{equation}
\label{eq:temp_condition}
\frac{2m\varkappa}{k} K\left(1/k\right)=\beta.
\end{equation}

The second condition (\ref{eq:integral_cond2}) can be written in the form of
\begin{eqnarray}
\label{eq:integral_cond2a}
\frac{1}{2\beta}\int_{0}^{2\beta}y_{m}(\prt)d\prt=\frac{m}{\beta}\int_{b}^{c}\frac{ydy}{\sqrt{-2U(y)}}= \nonumber \\
\frac{m\varkappa}{\beta}\int_{-\frac{1}{k}}^{\frac{1}{k}}\frac{-B+Dz}{A-Cz}\frac{dz}{\sqrt{(1-z^{2})(1-k^{2}z^{2})}}=  \\
\frac{m\varkappa}{\beta}\int_{0}^{\frac{1}{k}}\frac{2}{AC}\left(\frac{AD-BC}{1-\frac{C^{2}}{A^{2}}z^{2}}-DA\right)\frac{dz}
{\sqrt{(1-z^{2})(1-k^{2}z^{2})}}=\nonumber\\
\frac{2m\varkappa D}{\beta C}\left[\left(1-\frac{BC}{AD}\right)\Pi\left(\frac{C^{2}}{A^{2}},\frac{1}{k},k\right)-F\left(\frac{1}{k},k\right)\right].\nonumber
\end{eqnarray}
Here $\Pi(n,x,m)$ is the incomplete elliptic integral of the third kind \cite{Abramowitz64} defined as
\begin{equation}
\label{eq:elliptic_Pi}
\Pi (n,x,k)=\intop_{0}^{x}\frac{dz}{\left(1-nz^2\right)\sqrt{\left(1-z^2\right)
\left(1-k^2z^2\right)}}.
\end{equation}
The third integral condition (\ref{eq:integral_cond3}) is given by replacing $A$ with $B$ and $C$ with $D$ in \eref{eq:integral_cond2a},
\begin{eqnarray}
\frac{1}{2\beta}\int_{0}^{2\beta}y_{m}^{-1}(\prt)d\prt=\frac{m}{\beta}\int_{b}^{c}\frac{y^{-1}dy}{\sqrt{-2U(y)}}= \nonumber \\
\frac{m\varkappa}{\beta}\int_{-\frac{1}{k}}^{\frac{1}{k}}\frac{-A+Cz}{B-Dz}\frac{dz}{\sqrt{(1-z^{2})(1-k^{2}z^{2})}}=  \\
\frac{2m\varkappa C}{\beta D}\left[\left(1-\frac{AD}{BC}\right)\Pi\left(\frac{D^{2}}{B^{2}},\frac{1}{k},k\right)-F\left(\frac{1}{k},k\right)\right]. \nonumber
\end{eqnarray}
As in \eref{eq:complete_elliptic}, these results can be written in terms of a
complete elliptic integral using the transformation $w=kz$,
\begin{eqnarray}
\Pi (n,1/k,k) &=& \intop_{0}^{1/k}\frac{dz}{\left(1-nz^2\right)\sqrt{\left(1-z^2\right)\left(1-k^2z^2\right)}} \nonumber \\
& = & \intop_{0}^{1}\frac{dw/k}{\left(1-\frac{n w^2}{k^2}\right)\sqrt{\left(1-\frac{w^2}{k^2}\right)\left(1-w^2\right)}} =
 \frac{1}{k}\Pi (\frac{n}{k^2},\frac{1}{k}).
\end{eqnarray}
Here $\Pi (n,k)$ is the complete elliptic integral of the third kind.
The two integral conditions (\ref{eq:integral_cond2}),  (\ref{eq:integral_cond3}) are thus given by
\begin{equation}
\label{eq:ellipctic_cond2}
\frac{D}{C}\left[ 2\frac{1-BC/AD}{\beta  k / m\varkappa}
\Pi\left(\frac{C^2}{k^2A^2} ,\frac{1}{k}\right)-1\right]=r_C,
\end{equation}
\begin{equation}
\label{eq:ellipctic_cond3}
\frac{C}{D}\left[ 2\frac{1-AD/BC}{\beta  k / m\varkappa}
\Pi\left(\frac{D^2}{k^2B^2},\frac{1}{k}\right)-1\right]=\frac{r_B-r_A}{Q_C}.
\end{equation}

In order to obtain $K(\beta,r_\alpha),Q_A(\beta,r_\alpha)$ and $Q_C(\beta,r_\alpha)$
one has to invert \eref{eq:temp_condition},(\ref{eq:ellipctic_cond2}) and
(\ref{eq:ellipctic_cond3}). Once this is done the profile can computed by inverting
\eref{eq:HyberbplicTime}. The result may be expressed using the Jacobi elliptic
 function, $\mathrm{sn}\left(x,k\right)$,
defined by the equation $F\left(\mathrm{sn}\left(x,k\right),k\right)=x$.
The profile of $C$ particles is then given up to translations of $x$ by
\begin{equation}
\label{eq:analytic_profile_original}
\rho_C(x)=\frac{-B+D\mathrm{sn}\left(2\beta x / \varkappa,k \right) }
{A -C \mathrm{sn}\left(2\beta x / \varkappa,k \right)}.
\end{equation}
Note that the dependence on $m$ is hidden in $k,\varkappa,\alpha_{\pm}$.
The two other profiles, $\rho_A(x)$ and $\rho_B(x)$, are obtained from \eref{eq:rhoAofrhoC} and
$\rho_A(x)+\rho_B(x)=1-\rho_C(x)$.

\subsection{Two nonequal densities}

For convenience we write explicitly the solution for the two nonequal densities case, which is studied
extensively in this paper. The form of the solution in this case is very similar to that obtained
for equal densities in \cite{Ayyer2009}.

When $r_A=r_B\neq r_C$ it easy to see from symmetry that $Q_C=J_C/\beta=0$.
As a result, our solution depends only on two parameters $K,Q_A$ which are set
by $\beta,\Delta$. The latter is defined by
\begin{equation}
r_A=r_B=1/3-\Delta,\qquad r_C=1/3+2\Delta.
\end{equation}
The effective quartic potential which governs the motion of the particle is of the form
\begin{equation}	
U_{K,Q_{A},0}\left(y\right)=-\frac{1}{8}y^{2}(1-y)^{2}+\frac{K}{2}y-\frac{Q_{A}}{2}y^{2}
\end{equation}
and its four roots are thus $\{0,b,c,d\}$.
The M{\"o}bius transformation from these roots to the poles of the elliptic integral are obtained
by setting $a=0$ in the form presented in the previous section. More explicitly it is
given by
\begin{eqnarray}
\label{eq:mobius}
z=f\left(y\right)=\frac{\alpha_+y-1}{\alpha_-y+1},
\end{eqnarray}
where
\begin{equation}
\label{eq:apm_b}
\alpha_{\pm}=\frac{\pm bc+\sqrt{bc\left(d-c\right)\left(d-b\right)}}{bcd},
\end{equation}
and
\begin{equation}
\label{eq:small_k}
k=\frac{1+\alpha_{-}b}{1-\alpha_{+}b}.
\end{equation}
Note that this form is identical to the one defined in the equal densities case \cite{Ayyer2009}.

We now express using elliptic integrals the two conditions that determine mapping between $K,Q_A$ and $\beta,\Delta$.
The first condition is identical to \eref{eq:temp_condition},
\begin{equation}
\label{eq:temp_condition1}
2m\varkappa K\left(1/k\right)/k=\beta.
\end{equation}
where here $\varkappa$ takes the simpler form of
\begin{equation}
\label{eq:varkappa_b}
\varkappa = \frac{2(\alpha_{+}+\alpha_{-})}
{\sqrt{(1-\alpha_{+}b)(1-\alpha_{+}c)(1-\alpha_{+}d)}}.
\end{equation}
The second condition (\ref{eq:integral_cond2}) is given in this case as
\begin{equation}
\label{eq:r_condition}
\frac{1}{\alpha_{-}}\left[ 2\frac{1+\alpha_-/\alpha_+ }{\beta k /\varkappa m}
\Pi\left(\frac{\alpha_{-}^2}{\alpha_{+}^2k^2} ,\frac{1}{k}\right)-1\right]=r_C=\frac{1}{3}+2\Delta.
\end{equation}

The functions $K(\beta,\Delta),Q_A(\beta,\Delta)$ are obtained by inverting \eref{eq:temp_condition1} and \eref{eq:r_condition}. As in the previous section, we use
the result to express the profile of the $C$ particles,
\begin{equation}
\label{eq:analytic_profile_original_a}
\rho_C(x)=\frac{1+\mathrm{sn}\left(2\beta x / \varkappa,k \right) }
{\alpha_+ - \alpha_- \mathrm{sn}\left(2\beta x / \varkappa,k \right)}.
\end{equation}
The two other profiles, $\rho_A(x)$ and $\rho_B(x)$, are again obtained from \eref{eq:rhoAofrhoC} and
$\rho_A(x)+\rho_B(x)=1-\rho_C(x)$.

\section{Asymptotic behaviour at low temperatures}
\label{sec:low_temp}

In this section we study the behaviour of the hydrodynamic solution of model at
 low temperature ($L\to \infty$, $1 \ll \beta \ll L$) for the case
of two nonequal densities, $r_A=r_B\neq r_C$. This form will be especially useful in
future studies of a generalized ABC model with nonconserving dynamics \cite{Cohen2011}.
The $T=0$ limit corresponds to $K=Q_A=0$. Inserting this into
 \eref{eq:apm_b},\eref{eq:small_k} and \eref{eq:varkappa_b} we obtain
\begin{equation}
k = 1,\qquad \alpha_\pm = 1, \qquad \varkappa = 4.
\end{equation}
For $k=1$ the elliptic integral in \eref{eq:temp_condition1} diverges, corresponding to the limit of $\beta\to\infty$.
The elliptic integral in \eref{eq:r_condition} diverges as well, maintaining a finite value of $r_C$.

To study the low temperatures behaviour we assume that $K(\beta,\Delta)$ and $Q_A(\beta,\Delta)$ vanish for exponentially to leading order as
\begin{equation}
K\sim e^{-\beta \gamma_1}, \qquad Q_A\sim e^{-\beta \gamma_2}.
\end{equation}
In order to find the coefficients $\gamma_1$ and $\gamma_2$ we first examine the behaviour of \eref{eq:temp_condition1}.

Here we analyze only the behaviour of the $m=1$ solution, which is considered to be
the ground state of the model (see \sref{sec:phase_diagrams}).
We expand $\varkappa$ and $k$ around their value at $T=0$ as
\begin{equation}
\varkappa = 4 + \varkappa_1(K,Q_A) \qquad k = 1 + k_1(K,Q_A)
\end{equation}
where $\varkappa_1$ and $k_1$ are functions whose form is not written explicitly in order
to avoid lengthy expressions. They obey $0<\varkappa_1\ll 1$ and $0<k_1\ll 1$.
 Expanding \eref{eq:temp_condition1} to leading
order in these functions \cite{Mathematica8} we find that
\begin{equation}
\label{eq:k1}
\ln( k_1 ) = - \beta / 4 + O(1)
\end{equation}
The condition coming from the average density (\ref{eq:r_condition}) involves
an additional function,
\begin{equation}
\alpha_-/\alpha_+ = 1 - a_1(K,Q_A)
\end{equation}
which can be shown to obey $k_1\ll a_1 \ll 1$.
Expanding \eref{eq:r_condition}
in $a_1$ and $k_1$ to leading order \cite{Mathematica8} we obtain that
\begin{equation}
\label{eq:a1}
\ln(k_1) - 3 \ln(a_1) = 12\Delta \ln(a_1) + O(1).
\end{equation}
 Solving \eref{eq:k1} and \eref{eq:a1} to lowest order in $K$ and $Q_A$
yields
\begin{equation}
\label{eq:scale}
K\sim e^{-\beta \left( 1/3-\Delta \right)}, \qquad Q_A \sim \left\{ \begin{array}{ccc}
e^{-\beta\left(1/3-\Delta\right)} & \quad & \Delta\geq0\\
e^{-\beta\left(1/3+2\Delta\right)} & \quad & \Delta<0\end{array}\right. ,
\end{equation}
which agrees with form of $K\sim e^{-\beta/3}$ found in the equal densities case \cite{Barton2011}.

In order to interpret this result we observe the behaviour of \eref{eq:KDef}
whereby
\begin{equation}
\rho_{A}(x)\rho_{B}(x)\rho_{C}(x)=K-Q_{A}\,\rho_{C}(x),
\end{equation}
and hence
\begin{equation}
\int_0^1 dx \rho_A(x)\rho_B(x)\rho_C(x) \sim e^{-\beta \min(\frac{1}{3}-\Delta,\frac{1}{3}+2\Delta)}
= q^{\min(N_A,N_B,N_C)}.
\end{equation}
In the limit of $\beta \to \infty$ the number of triplets of $ABC$ is governed by the probability of an event
where in the fully separated state two particles of different species meet in the domain of the third
species. This probability scales as $q^{\min(N_A,N_B,N_C)}$, because this
 event occurs with the highest probability in the smallest domain.

\bibliographystyle{unsrt}
\bibliography{abc_sol}

\begin{thebibliography}{10}

\bibitem{Spohn1983}
H~Spohn.
\newblock Long range correlations for stochastic lattice gases in a
  non-equilibrium steady state.
\newblock {\em J. Phys. A}, 16:4275--4291, 1983.

\bibitem{Garrido1990}
P~L Garrido, J~L Lebowitz, C~Maes, and H~Spohn.
\newblock Long-range correlations for conservative dynamics.
\newblock {\em Phys. Rev. A}, 42(4):1954--1968, 1990.

\bibitem{Dorfman1994}
J~R Dorfman, T~R Kirkpatrick, and J~V Sengers.
\newblock Generic long-range correlations in molecular fluids.
\newblock {\em Annu. Rev. Phys. Chem.}, 45(1):213--239, 1994.

\bibitem{Ortiz2004}
J~M Ortiz~de Z{\'a}rate and J~V Sengers.
\newblock On the physical origin of long-ranged fluctuations in fluids in
  thermal nonequilibrium states.
\newblock {\em J. Stat. Phys.}, 115:1341--1359, 2004.

\bibitem{Sadhu2011}
T~Sadhu, S~N Majumdar, and D~Mukamel.
\newblock Long-range steady state density profiles induced by localized drive.
\newblock {\em ArXiv e-prints}, arXiv:1106.1838, 2011.

\bibitem{Evans1995}
M~R Evans, D~P Foster, C~Godr\`eche, and D~Mukamel.
\newblock Spontaneous symmetry breaking in a one dimensional driven diffusive
  system.
\newblock {\em Phys. Rev. Lett.}, 74(2):208--211, 1995.

\bibitem{Lahiri1997}
R~Lahiri and S~Ramaswamy.
\newblock Are steadily moving crystals unstable?
\newblock {\em Phys. Rev. Lett.}, 79(6):1150--1153, 1997.

\bibitem{Lahiri2000}
R~Lahiri, M~Barma, and S~Ramaswamy.
\newblock Strong phase separation in a model of sedimenting lattices.
\newblock {\em Phys. Rev. E}, 61(2):1648--1658, 2000.

\bibitem{Evans1998}
M~R Evans, Y~Kafri, H~M Koduvely, and D~Mukamel.
\newblock Phase separation in one-dimensional driven diffusive systems.
\newblock {\em Phys. Rev. Lett.}, 80(3):425--429, 1998.

\bibitem{Evans1998b}
M~R Evans, Y~Kafri, H~M Koduvely, and D~Mukamel.
\newblock Phase separation and coarsening in one-dimensional driven diffusive
  systems: Local dynamics leading to long-range hamiltonians.
\newblock {\em Phys. Rev. E}, 58:2764--2778, 1998.

\bibitem{Clincy2003}
M~Clincy, B~Derrida, and M~R Evans.
\newblock Phase transition in the abc model.
\newblock {\em Phys. Rev. E}, 67:066115, 2003.

\bibitem{Bodineau2008}
T~Bodineau, B~Derrida, V~Lecomte, and F~van Wijland.
\newblock Long range correlations and phase transitions in non-equilibrium
  diffusive systems.
\newblock {\em J. Stat. Phys.}, 133:1013--1031, 2008.

\bibitem{Lederhendler2010a}
A~Lederhendler and D~Mukamel.
\newblock Long-range correlations and ensemble inequivalence in a generalized
  $abc$ model.
\newblock {\em Phys. Rev. Lett.}, 105(15):150602, 2010.

\bibitem{Lederhendler2010b}
A~Lederhendler, O~Cohen, and D~Mukamel.
\newblock Phase diagram of the abc model with nonconserving processes.
\newblock {\em J. Stat. Mech: Theory Exp.}, 2010(11):P11016, 2010.

\bibitem{Cohen2011}
O~Cohen and D~Mukamel.
\newblock to be published.

\bibitem{Kac1963}
M~Kac, G~E Uhlenbeck, and P~C Hemmer.
\newblock On the van der waals theory of the vapor-liquid equilibrium.
\newblock {\em J. Math. Phys.}, 4(2):216--228, 1963.

\bibitem{Ayyer2009}
A~Ayyer, E~A Carlen, J~L Lebowitz, P~K Mohanty, D~Mukamel, and E~R Speer.
\newblock Phase diagram of the abc model on an interval.
\newblock {\em J. Stat. Phys.}, 137(5-6):1166--1204, 2009.

\bibitem{Barton2010}
J~Barton, J~L Lebowitz, and E~R Speer.
\newblock The grand canonical abc model: a reflection asymmetric mean-field
  potts model.
\newblock {\em J. Phys. A}, 44(6):065005, 2011.

\bibitem{Barton2011}
J~Barton, J~L Lebowitz, and E~R Speer.
\newblock Phase diagram of a generalized abc model on the interval.
\newblock {\em ArXiv e-prints}, arXiv:1106.1942, 2011.

\bibitem{Evans2000}
M~R Evans.
\newblock Phase transitions in one-dimensional nonequilibrium systems.
\newblock {\em Braz. J. Phys.}, 30:42--57, 2000.

\bibitem{Fayolle2004}
G~Fayolle and C~Furtlehner.
\newblock Stochastic deformations of sample paths of random walks and exclusion
  models.
\newblock In M~Drmota, P~Flajolet, D~Gardy, and B~Gittenberger, editors, {\em
  Mathematics and Computer Science III: Algorithms, Trees, Combinatorics and
  Probabilities (Trends in Mathematics)}, pages 415--427. Birkh\"auser, Basel,
  2004.

\bibitem{Abramowitz64}
M~Abramowitz and I~Stegun.
\newblock {\em Handbook of Mathematical Functions with Formulas, Graphs and
  Mathematical Tables}.
\newblock National Bureau of Standards Appl. Math. Series, 1964.

\bibitem{Mathematica8}
Inc. Wolfram~Research.
\newblock {\em Mathematica Edition: Version 8.0}.
\newblock Wolfram Research, Inc., Champaign, Illinois, 2010.

\end{thebibliography}

\end{document}